
\input phyzzx

\nopubblock
\PHYSREV
\parindent=0 truecm
\parskip=0 truecm
\hsize=160mm
\baselineskip=12pt

\titlepage
\title{ASTROPHYSICAL EVIDENCE ON PHYSICS BEYOND THE STANDARD
MODEL\foot{Invited talk to be published in ``Future Physics
and Accelerators'' (ed. M. Chaichian), proceedings
of the 1st Arctic Workshop on Future Physics And Accelerators,
Sariselka, Lapland, Finland, August 16-22, 1994.}}
\author{Arnon Dar\foot{Supported in part by the Technion Fund
for Promotion of Research }
\address{Department of Physics and Space Research Institute,
Technion - Israel Institute of Technology, Haifa 32000, Israel}}
\abstract
Astrophysics and cosmology can be used to test the standard model
of particle physics under conditions and over
distance and time scales not accessible to laboratory experiments.
Most of the astrophysical observations are in good agreement
with the standard model. In particular, primordial nucleosynthesis,
supernova explosions, stellar evolution and cosmic background
radiations have been used to derive strong limits on physics
beyond the  standard model.
However, the solution of some important astrophysical and cosmological
problems may require new physics beyond the standard
model. These include the origin of the initial conditions,
large scale structure formation, the baryon asymmetry
in the observed Universe, the dark matter problem, the solar
neutrino problem and some cosmic ray puzzles. Here I review  some
important developments relevant to some of
these problems, which took place most recently.
\endpage
\hsize=160mm
\baselineskip=12pt
Gravitational lensing observations have become a very powerful tool for
veryfying the existence of dark matter, for mapping its distribution
in galactic halos and clusters and for studying its nature. Here we
consider two important  applications of gravitational lensing in
search of new physics.
\smallskip
{\bf 1. TESTS OF GENERAL RELATIVITY AT LARGE DISTANCES AND DARK
MATTER}
\smallskip
There are at least two good reasons for testing the validity of
Einstein's General Relativity (EGR) and its weak field limit,
Newtonian Gravity (NG), over cosmic distances;

(a) It has not been tested before over such distances -
 All astronomical tests of EGR and NG, so far, were limited to
the solar system (see e.g., Weinberg 1972; Will 1984)
 and to close binary systems (Damour and Taylor 1991)
(PSR 1913+16, 4U1820-30 and PSR 0655+64), i.e.,
to distance scales less than a few Astronomical Units,
whereas EGR and NG have been applied to astronomical systems such as
galaxies, clusters of galaxies, superclusters and the whole
Universe, which are typically $10^6-10^{15}$ times larger.

(b) The dark matter problem -
All the dynamical evidence from galaxies,
clusters of galaxies, superclusters and large scale structures
that they contain vast quantities of non luminous dark matter
(for a recent review see e.g., Gould 1995) has been obtained assuming
the validity of EGR or Newton's laws
for such systems. As we shall see below
there are good reasons to believe that most of this dark matter
is non baryonic. However,
in spite of extensive laboratory searches
no conclusive evidence has been found either
for finite neutrino masses (minimal extension of the standard model)
or for the existence of other
particles beyond the standard model that can form
dark matter. This
has led some authors to question the validity of EGR and NG
over large distances and to suggest (see for instance Sanders 1990
and references therein) that perhaps EGR and NG
are only approximate theories of gravity and that a correct
theory of gravity will eliminate the dark matter problem. Indeed
alternative theories (Mannheim and Kazanas 1989)
to General Relativity or modifications of Newton's laws
(e.g., Milgrom and Beckenstein 1984;
Sanders 1984) have been proposed
in order to explain the observations without invoking dark matter.
\smallskip
Gravitational macro lensing observations can be used to test EGR
and NG over cosmological distances (Dar 1992;
Dar 1993). EGR predicts that light which
passes at an impact parameter $b$ from a spherical symmetric mass
distribution is deflected
by an angle which, for small angles, is given approximately by
$$ \alpha\approx {4GM(b)\over c^2b}~, \eqno \eq$$
where G is Newton's gravitational constant and $M(b)$ is the
mass interior to $b$.  The mass $M(r)$ enclosed within a radial distance
$r$ from the center is given by Kepler's third law
$ M(r)\approx v_{cir}^2r/ G~, $
where $v_{cir}$ is the circular velocity of a mass orbiting at
a distance $r$ from the center. Consequently, spiral
galaxies, which have flat rotation curves $(v_{cir}\approx const.)$
have $M(r)\propto r$, $\rho(r)\propto
1/r^2~,$  and $M(b)/b \approx \pi v_{cir}^2/2G~, $ which give
rise to a constant deflection angle independent of impact parameter,
$$ \alpha=2\pi \left ({v_{cir}\over c}\right )^2~.  \eqno\eq $$
For large spiral galaxies,
$v_{cir}\sim 250~km~s^{-1}$ and $\alpha\sim 1"~.$  In elliptical
galaxies, or clusters of galaxies, whose total mass distributions are
well described by singular isothermal sphere distributions
$\rho(r)\approx (1/2\pi G)(\sigma_{_\parallel}/c)^2r^{-2}~,$ the squared
circular velocity is replaced by $v_{cir}^2=2\sigma_{_\parallel}^2~,$
where $\sigma_{_\parallel}$ is the one-dimensional line-of-sight
velocity dispersion in the galaxy or the cluster, respectively.
For a typical large elliptical galaxy with $\sigma_{_\parallel}\sim
200~km~s^{-1}$ the constant deflection angle is $\alpha\sim 1.5"$
while for a rich cluster with $\sigma_{_\parallel} \sim 1000~km~s^{-1}$
the constant deflection angle is $\alpha\sim 30".$ Hence, the
Large optical telescopes, VLA and VLBI radio telescopes
have been used to discover and study gravitational lensing of
quasars and galaxies by galaxies and clusters of galaxies (see , e.g.,
Blandford and Narayan 1992).

EGR and Newton's laws can be tested over
galactic and intergalactic distances by comparing the deflection
of light which is extracted from these observations
and the deflection of light which is predicted from the measured
rotation curves or line-of-sight velocity dispersions in these systems.
However the deflection of light cannot be measured directly and
must be deduced from the multiple image pattern (angular positions and
relative magnifications) of the source which is produced by the lens.
Generally, this requires a complicated inversion procedure
(see e.g. Blandford and Narayan  1992 and references therein)
and additional assumptions. However, for testing EGR
one can select the gravitational lensing cases where the
lens is simple, the pattern-recognition is straightforward
and the deflection angle can be read directly
from the simple multiple image pattern:
\smallskip
{\bf Einstein Rings, Crosses, and Arcs}:
On the rare occasion that a lensing galaxy with a radially symmetric
surface density happens to lie on the line-of-sight to a distant quasar
it forms in the sky a ring image (Cholson 1924; Einstein 1936) of the
quasar around the center of the lensing galaxy, whose angular diameter is
$$\Delta \theta=4\theta_r\approx 2{D_{LS}\over D_{OS}}\alpha
\approx 4\pi {D_{LS}\over D_{OS}}\left({\sigma_{_\parallel}\over c}
\right)^2~,\eqno\eq $$
where $D_{OL}$, $D_{LS}$, and $D_{OS}$ are the Obsever-Lens,
Lens-Source and Observer-Source angular diameter distances, respectively.
Five Einstein rings,
MG1131+0456, 1830-211, MG1634+1346, 0218+357 and MG1549+3047
were discovered thus far by high resolution radio observations (see
e.g., Blandford and Narayan 1992 and references therein)
but, only for MG1634+1346 are the redshifts of both the lens and the
ring image known, allowing a quantitative test of EGR.
When the source is slightly off center, the ring breaks into a pair of
arcs, as actually observed for the ring image MG1634+1346 of a radio
lobe of a distant quasar (Langston et al 1989, 1990).
\smallskip
When the lens has an elliptical surface density and the
line of sight to the source passes very near its center,
the Einstein ring degrades into an ``Einstein Cross'', i.e.,
four images that are located
symmetrically along the two principal axes
(and a faint fifth image at the center), with a mean angular
separation between opposite images given approximately
by Eq.3, as observed in the case of Q2237+0305
(Rix et al 1992 and references therein).

\smallskip
When an extended distant source, such as a galaxy, lies on a cusp
caustic behind a giant elliptical lens, such as a rich cluster
of galaxies, it appears as an extended luminous arc
on the opposite side of the lens (Grossman and Narayan 1988,
Blandford et al 1989). The angular distance
of the arc from the center of the lens
is given approximately by the radius of the Einstein ring.
Giant arcs (Soucail et al 1988; Lynds and Petrosian 1989)
were discovered, thus far, in the central regions of 13
rich clusters (see e.g., Blandford and Narayan 1992 and references
therein) and in six cases, Abell 370, 963 and 2390, Cl0500-24,
Cl2244-02 and Cl0024+1654 the redshifts of both the giant arc image
and the cluster are known and the velocity dispersion in the cluster
has been estimated from the redshifts of the member galaxies or the
X ray emission, allowing a quantitative test of EGR.
\smallskip
{\bf Gravitational Time Delay}:
In the thin lens approximation the time delay predicted by EGR
is a sum of the time delay due to the difference in
path length between deflected and undeflected light rays
and the time delay due to the different
gravitational potential felt by the light rays
$$ \Delta t\approx (1+z_{_L})\left[{D_{OL}D_{OS}\over 2c D_{LS}}
(\vec{\theta}_{_I}-\vec{\theta}_{_S})^2~- {\phi(\theta_{_I})
\over c^3}\right]~, \eqno\eq $$
where $\phi(\theta_{_I})$ is the gravitational potential
of the lens at $\theta_{I}$. Thus, the time delay between two images A,B,
due to a lensing galaxy with nearly spherical isothermal mass distibution
that lies near the line-of-sight to the source
(even if it is embedded
in a large cluster with an approximately constant deflection angle
over the whole image), in the thin lens approximation
reduces to a simple form,
$$\Delta t_{A,B}\approx 2\pi(1+z_{_L})~
(\vert\vec{\theta}_A\vert -\vert\vec{\theta}_B\vert)\left({
\sigma_{_\parallel}\over c}
\right)^2 {D_{OL}\over c}~, \eqno\eq $$
which can also be written as $$\Delta t_{A,B}\approx (1+z_{_L})
(\vert\vec{\theta}_A\vert -\vert\vec{\theta}_B\vert)
\vert\vec{\theta}_A -\vec{\theta}_B\vert {D_{OS} \over D_{LS}}
 {D_{OL}\over 4c}~. \eqno\eq $$
Note that while the deflection angle is dimensionless, i.e.,
depends only on dimensionless parameters, the time delay is
dimensionfull and depends on the absolute value of the Hubble
parameter (through $D_{OL}$) and can be used to measure $H_0$.
For the double quasar Q0957+561 Eq. 6 yields $H_0\approx (76\pm4)/
\Delta t_(Y)\approx 70\pm5~km~s^{-1}Mpc^{-1}$, for $\Delta t\approx
1.1Y$  (Vanderriest et al 1989; Schild 1990; Pelt et al 1994).

\smallskip
Fig. 1 summarizes our comparison between the above
EGR predictions and  observations on the most simple known
cases of gravitational lensing of quasars and galaxies
by galaxies or cluster of galaxies. These include
the Einstein Ring MG1654+1346, the Einstein Cross Q2237+0305,
the giant Einstein Arcs in the clusters
A370, Cl2244-02 and  Cl0024+1654 and
the time delay between the two images of the Quasar Q0957+561.
The agreement
between the predictions and the observations confirm within errors
(the error bars are statistical
only and do not include model uncertainties)
the validity of EGR and NG over distances of 0.1kpc - 0.1 Mpc,
i.e., $\sim 10^6-10^9$ times larger than the size of the solar system.
Moreover, the gravitational lensing observations  confirm
that most of the mass of galaxies, groups and clusters of galaxies
consists of dark matter and (following Tyson et al 1990 and
Tyson 1991) have been used
extensively to map the distribution of dark matter in clusters
of galaxies.
\bigskip
{\bf 2. EVIDENCE FOR  NON BARYONIC DARK MATTER}.
\smallskip
The astrophysical evidence for non baryonic dark matter
is considered by many to be
the best evidence for physics beyond the standard model.
The main evidence for non baryonic dark matter  comes from comparisons
between the average  densities of baryonic and gravitating matter in the
Universe (e.g. Kolb and Turner 1991). The average baryon density is
best inferred from Big-Bang Nucleosynthesis, while the average
density of gravitating matter in the Universe is presently
best determined from the dynamics of clusters of galaxies and large
scale structures, from X-ray observations of clusters and from
gravitational lensing observations. We shall first review this evidence.
\smallskip
{\bf 2.1  The Mean Baryon Density From Big-Bang Nucleosynthesis}
\smallskip
The predictions of the Standard Big-Bang Nucleosynthesis (SBBN) theory
(Peebles 1966; Wagoner, Fowler and Hoyle 1967; Wagoner 1973; Yang et al
1984) for the primordial abundances of the light elements (H, D, $^3$He,
$^4$He, and $^7$Li) depend on low energy nuclear cross sections and on
three additional parameters, the number of flavours of light neutrinos,
${\rm N_\nu},$the neutron lifetime, ${\rm \tau_n~}$ and the ratio
of baryons to photons in the Universe, ${\rm \eta\equiv n_b/n_\gamma~}$.
The relevant nuclear cross sections are known from laboratory
measurements (e.g., Caughlan and Fowler 1988 and references therein).
The measurements at the Large Electron Positron Collider (LEP) at CERN
gave ${\rm N_\nu=3.04\pm 0.04~}$ (e.g., Mana and Martinez 1993).
Measurements of $\tau_n$ in neutron bottles and Penning traps coupled
with previous measurements yielded the weighted average (see Particle
Data Group 1994) $\tau_n=887\pm 2.0~s~. $ Finally, measurements of
the cosmic microwave background radiation by COBE (Mather et al 1994)
gave a black body temperature  ${\rm T=2.726\pm 0.017~K},$
which yields ${\rm n_\gamma=20.28T^3\approx 411\pm 8~cm^{-3}}. $ Hence,
SBBN theory predicts quite accurately (see Fig. 1) the primordial
abundances of the light elements as function of a single unknown
parameter, ${\rm n_b~},$ the mean baryon number density in the Universe.
Thus, the primordial abundances of the light elements, as inferred
from observations, can be used to test SBBN theory and determine
this number. Indeed, it has been claimed  repeatedly that the
predictions of SBBN theory agree with observations if
$\eta_{10}\equiv \eta\times 10^{10}\approx 4,$ which implies that most
of the nucleons in the Universe are dark (e.g., Kolb and Turner 1990;
Walker et al 1991; Smith et al 1993 and references therein). Moreover,
based on these analyses, many limits on physics beyond the standard
particle physics model (new interactions; new weakly interacting
particles; additional neutrino flavours; masses, mixings,
magnetic moments, decay modes and lifetimes of neutrinos) were
derived by various authors.
\smallskip
However, the claimed concordance between SBBN theory and the observed
abundances of the light elements extrapolated to their primordial values
had a rather poor confidence level, was demonstrated for primordial
abundance of  $^4$He which deviated significantly from its best value
as inferred from observations and relied heavily on the highly uncertain
extrapolated values for the primordial abundances of D+$^3$He.
Hence, SBBN could provide neither reliable evidence that most
of the baryons in the universe are dark nor reliable limits on
physics beyond the standard particle physics model (Dar, Goldberg
and Rudzsky 1992). In fact  Dar, Goldberg and Rudzsky (1992) argued
that the theoretical upper bound on primordial D+$^3$He that
was estimated by Walker et al (1991) from a gallactic evolution model
is highly uncertain and the best values of the primordial
abundances of $^4$He, and $^7$Li as inferred from observations
indicate that $\eta_{10}\approx 1.60\pm 0.10~$. This value yields a
mean baryon mass density in the present Universe which is not
significantly larger than the total mass density of matter visible
in the V, IR, UV, X and Radio bands, provided  that the true value of
the Hubble constant is close to its value measured recently (Freedman
et al 1994) by the
repaired Hubble Space Telescope, $H_0=80\pm 17~km~s^{-1}~Mpc^{-1}.$
It predicts, however,
a primordial abundance (by numbers) of D, ${\rm [D]_p/[H]_p
\approx (2.10\pm 0.20)\times 10^{-4}}$, which is larger by about
an order of magnitude than that observed in the galactic interstellar
medium (Linsky 1993).
\smallskip
During the past three years new observations and refined analyses
have greatly improved the estimated values of the primordial
abundances of the light elements:

{\bf Helium 4:}
The most accurate determinations
of the primordial abundance of $^4$He are based on measurements of
its recombination radiation in very low metallicity extragalactic
HII regions which are the least contaminated by stellar production of
$^4$He. A number of groups have obtained high-quality data for very
metal-poor, extragalactic HII regions which they
used to extrapolate to zero metallicity yielding a primordial abundance
(by mass) of $Y_p=0.228\pm 0.005 $ (Pagel et al 1992),
$Y_p=0.226\pm 0.005$ (Mathews et al 1992),
$Y_p=0.230\pm 0.005$ (Skillman and Kennicutt 1993),
$Y_p=0.229\pm 0.004$ (Izotov et al 1994),
where $1\sigma$ statistical
and systematic errors were added in quadrature. A weighted
average yields
$$Y_p=0.228\pm 0.005~. \eqno\eq $$
It is not inconceivable that systematic errors (e.g., due to collisional
excitation, contribution of neutral Helium, interstellar reddening,
UV ionizing radiation, grain depletion, non homogeneous density and
temperature, etc.) are larger;
however, there is no empirical evidence for that.

{\bf Deuterium:}
Deuterium is easily destroyed already at relatively low temperatures.
Consequently, its abundance observed today can only
provide a lower limit to the big-bang production.
Measurements of its abundance
in the local interstellar medium (LISM)
made recently by the Hubble Space Telescope (Linsky et al. 1993), gave
 [D]/[H]$ = (1.65^{+0.07}_{-0.18})\times 10^{-5}.$
{}From the analysis of solar-wind particles captured
in foils exposed on the moon and studies of primitive meteorites,
Geiss (1993) deduced a pre-solar abundance of
[D]/[H]$ = (2.6 \pm 1.0)\times 10^{-5}.$  These values can be used as
lower bounds on primordial Deuterium.
High redshift - low metallicity quasar absorption systems
offer the possibility of observing its
abundance back in the past in very primitive clouds (Webb et al 1991).
Recent measurements of the absorption  spectrum of the distant  quasar
Q0014+813 in a low-metallicity high redshift (z= 3.32)
hydrogen cloud, by Songaila et al
(1994) with the Keck 10m telescope at Mauna Kea, Hawaii, and
by Carswell et al (1994) with the 4m telescope at Kit Peak, Arizona
showed an absorption line at the expected position of the
isotopically shifted Lyman $\alpha$ line of Deuterium.
The line shape was best fitted with Deuterium abundance of
$${\rm [D]/[H] \approx  2.5 \times 10^{-4}}. \eqno\eq $$
(The probability that the absorption line is due to a second
hydrogen cloud with a Lyman $\alpha$ absorption line
at the position of the isotopically shifted deuterium line,
was estimated as 3\% and 15\% by Songaila et al (1994)
and Carswell et al (1994), respectively.)
The above value is an order of magnitude larger than the
interstellar value and a factor of three larger than the 95\%
confidence level upper bound on the primordial abundance
of D+$^3$He that was inferred by Walker et al (1991).
However, Walker et al (1991) used an uncertain galactic chemical
evolution model to extrapolate their estimated
presolar D+$^3$He abundance to zero cosmic age.
Moreover, interstellar measurements of D and $^3$He abundances
show large variations from site to site and the solar system values
may not be a typical sample of galactic material 4.5 GY ago.

{\bf Helium 3:} From measurements of $[^3$He]/$[^4$He] in meteorites
and the solar wind Geiss (1993) concluded that the presolar
abundance of $^3$He is $[^3$He]/[H]=$(1.5\pm 0.3)\times 10^{-5}$.
However, any further extrapolations to zero cosmic age
of the $^3$He (or the $^3$He+D) abundance extracted
from solar system or interstellar observations are highly
uncertain because
$^3$He is both produced (via D(p,$\gamma)^3$He) and destroyed
(via $^3$He($^3$He,2p)$^4$He and $^4$He($^3$He,$\gamma)^7$Be)
in early generation stars.
Hogan (1994) has recently suggested that
the envelope material in low mass stars is mixed
down to high temperature after they reach the giant branch, so that the
$^3$He is destroyed before the material is ejected.
Indeed from radio observations
of highly ionized Galactic HII regions Balser et al (1994)
and Wilson and Rood (1994) inferred  [$^3$He]/[H] values that
ranged between
 $(6.8\pm 1.5)\times 10^{-6}$ for W49 and
 $(4.22\pm 0.08)\times 10^{-5}$ for W3.
Hyperfine emission in the  planetary nebula N3242
indicates  (Rood, Bania and Wilson 1992)
a large enrichment, $^3$He/H$\approx 10^{-3}$.
These spread values show
that the presently observed $^3$He abundances apparently reflect
complicated local chemical evolution and do not allow
a reliable determination of the primordial $^3$He abundance from
presently observed solar or LISM abundances.

{\bf Lithium 7:} The primordial abundance of $^7$Li
was determined from the most metal poor, Population II halo stars.
Such stars, if sufficiently warm $(T\gsim 5500K)$,
have apparently not depleted their surface Lithium and are expected
to have nearly a constant $^7$Li  abundance reflecting its
abundance at the early evolution of the Galaxy (Spite and
Spite 1982a,b).
High-precision LiI observations of 90 extremely metal
poor halo dwarfs and main sequence turnoff stars
have been performed recently by Thorburn (1994).
{}From the surface $^7$Li abundances of the hottest metal-deficient
stars (${\rm T\sim 6400K})$ Thorburn estimated
$${\rm[^7Li]_p/[H]_p=(1.7\pm 0.4)\times 10^{-10}}.\eqno\eq $$
Thorburn's data suggest a slight
systematic variation of the $^7$Li abundance with surface
temperature, possibly indicating some depletion from a higher
primordial value by processes that transport $^7$Li inward to
regions where it can be burned.
However, the amount of depletion
is constrained by the relatively narrow spread in $^7$Li abundance
for a wide range of surface temperatures and metallicities and
by the observation of $^6$Li in population
II stars by Smith, Lambert, and Nissen (1993) and by
Thorburn (1994):
Big-bang production of $^6$Li is negligible. It is
presumably produced by cosmic-rays.
Since $^6$Li is burned much more easily than
$^7$Li and yet still observed with an abundance
expected for cosmic-ray production, depletion of $^7$Li
cannot have been very significant.
\smallskip
In Fig. 2 we compare the predictions of the
SBBN theory and the observed abundances of the light elements
extrapolated to their primordial values.
The confidence level of the agreement
between the two
using the standard $\chi^2$ test as function of
$\eta_{10}$ is also shown in Fig. 2.
(Errors were assumed to be statistical in nature. Experimental and
theoretical uncertainties were added in quadrature). Fig. 2 shows
that the primordial abundances of the light elements as inferred
from observations are in very good agreement
(confidence level higher than 70\%) with those predicted by SBBN theory
for $\eta_{10}\approx 1.60\pm 0.1~.$
The corresponding mean cosmic baryon number density is
${\rm n_b=\eta n_{\gamma}= (6.6\pm 0.5)\times
10^{- 8}cm^{-3}},$  which yields a baryon mass density
(in critical density units ${\rm \rho_c\equiv 3H_0^2/8\pi G~}$) of
$${\rm \Omega_b\equiv \rho_b/\rho_c=(0.0058\pm 0.0007)h^{-2}}\approx
                 0.01\pm 0.004~, \eqno\eq$$

where $h=0.80\pm 0.17$
is the Hubble constant in units of $100~km~s^{-1}Mpc^{-1}$
measured by the repaired Hubble Space Telescope (Freedman et al 1994).
\smallskip
{\bf 2.2 The Baryonic Mass Fraction in Clusters of Galaxies}
\smallskip
Rich clusters of galaxies are the largest objects for which
total masses can be estimated directly. In fact, the need for
astrophysical dark matter was first identified for such
systems by Zwicky in 1933.

The total mass enclosed within a distance R from the centers
of clusters of galaxies has been determined by three independent
methods:

a) From the virial theorem applied to the radial velocities
of cluster members assuming that the velocities are
distributed isotropically and that light traces mass.

b) From analyses of the distribution of giant arcs and arclets
produced by gravitational lensing of distant galaxies
by the gravitating mass in clusters of galaxies.

c) From the X-ray emission of intergalactic hot gas which is
trapped in the deep gravitational potential of rich clusters,
under the assumption that the gas is relaxed.
\smallskip
All three methods yield similar results. When coupled with
photometric measurements of the light emitted by the galaxies
in the clusters they yield
an average total mass to blue light ratio of $<M/L>=
(230\pm 30)hM_\odot/L_\odot.$ The density of blue light
in the Universe was measured (e.g., Loveday et al 1992)
to be  $\rho_L=(1.83 \pm 0.35)\times 10^8h^2Mpc^{-3}$.
If the mean M/L ratio for clusters  represents well the mean M/L
ratio in the Universe then the mean cosmic density is
$$\Omega={\rho_L<M/L>\over \rho_c}\approx 0.15h\approx0.12\pm 0.2~.
\eqno\eq $$
This density is larger
by more than an order of magnitude than the baryon density
inferred from Big-Bang Nucleosynthesis and provides the best
evidence for non baryonic dark matter. This conclusion is further
confirmed by recent observations with the ROSAT  X-ray Telescope:
The positional sensitive proportional counter (PSPC)
on board the ROSAT observatory has been used recently to
measure  the specral and spatial distribution of X-ray
emission from many rich clusters. These measurements have been
used to determine the total gravitating mass, $M_t$, of the clusters
and the fraction of that mass which is in the form of X-ray
emitting hot gas, $M_{gas}$. Photometric measurements of the light
emitted by the galaxies in the clusters have been used to estimate
the total stellar mass, $M_*$, in the clusters. It was found (e.g.,
Briel et al 1992, White et al 1993 and references therein) that
$M_*/M_t\approx 0.01$  and  $<M_{gas}/M_t>\approx 0.05h^{-3/2}$,
i.e., the known forms of baryonic matter account only for
a small fraction of the total mass. In fact, numerical simulations
of structure formation indicate that the ratio of baryonic to non
baryonic mass is preserved in cluster formation (e.g. White et al
1993). Consequently,
the observed baryonic fraction in clusters and Big-Bang
nucleosynthesis imply that
$$ \Omega\approx {M_t\over M_b}\Omega_b\approx {0.0058h^{-2}\over
   0.01+0.05h^{-3/2}}\approx 0.12\pm 0.2~, \eqno\eq $$
in good agreement with the above independent estimate.
If the cosmic dark matter consists of massive neutrinos then
the  neutrino masses satisfy $\Omega h^2\approx\Sigma m_\nu$, i.e.,
$\Sigma m_\nu \approx 7\pm 2 ~eV$. This is also the neutrino mass
which generates in a self consistent way (Tremaine and Gunn 1979)
the gravitational potentials and the sizes of clusters of galaxies
as determined from X-ray measurements and from the dispersion
of velocities of the galaxies in the clusters.
\smallskip
{\bf 2.3 Galactic Dark Matter and Gravitational Microlensing}
\smallskip
The observed flat rotation curves of spiral galaxies, including
our Milky Way (MW), indicate that they  have extensive halos
of dark matter (see e.g., Gould 1995). Paczynski (1986) has suggested
that if the halo dark matter is made of brown dwarfs (low mass
stars whose mass is below that required to ignite hydrogen, i.e.,
less than $0.08M_\odot$ for primordial chemical composition)
it can be detected by their gravitational lensing of background
stars. For galactic distances the splitting of the source into
multiple images is too small (typically micro arcsec)
to be resolved, but the lensing causes a typical
magnification of the source which is time dependent due to
the relative motion of the Earth, lens and source:
$$ A(t)= {u^2+2\over u\sqrt{u^2+4}};~~~ u(t)=\left[u_{min}^2
+\left({2(t-t_{max})\over \bar t}\right)^2\right]^{1/2}, \eqno\eq $$
where $\bar t=2r_{E}/v$ is the time for the line of sight
to move through two Einstein radii
$D_E=2r_E=4\sqrt{GMD_{OL}D_{LS}/D_{OS}}$
and $u(t)$ is the distance between the lens and the line of
sight in units of $r_{E}$.

Three experiments (MACHO, EROS and OGLE) reported (Alcock et al 1993;
Aubourg et al 1993; Udalski et al 1993; Sutherland et al 1995;
Moscoso 1995) the detection
of more than 50 microlensing events, most of which are in the
direction of the galactic center and only 5 are of stars in the
Large Magelanic Cloud (LMC) at a distance of $D_{OS}\approx 50~ kpc$.
The number of events both in the directions of the galactic center and of
the LMC is much more than expected from the known population of stars
in the MW, but the number of events in the direction of the LMC is
much smaller than expected if the MW halo is spherical and consists
entirely of MAssive Compact Halo Objects (MACHOs). In particular,
the MACHO experiment detected 3 microlensing events in $\sim 10^7$
star-year monitoring of LMC stars. They explored a range of different
halo profiles (Alcock et al 1995) and found
a total mass of MACHOs interior to 50 $kpc$
of $8^{+14}_{-6}10^{10}M_\odot$ relatively independent of the assumed
model for the MW halo. For a naive spherical halo model it implies
that the halo mass fraction in MACHOs is $f=0.2^{+0.33}_{-0.14}$
and the most likely MACHO mass is $M_L=0.06^{+0.11}_{-0.04}M_\odot,$
as demonstrated in Fig. 3.
\smallskip
Recently, Sackett et al 1994 reported the discovery of a faint red halo
in the edge-on galaxy NGC5907  with a radial density distribution
similar to that observed for dark halos  ($\rho\sim r^{-2}$)
and an inferred M/L ratio of 350-500, similar to that expected for
subdwarf stars. This suggested that its dark halo is actually a faint
red halo made of subdwarf stars.
As NGC5907 is similar to our MW in type and radius it also suggests
that subdwarf stars constitute the MW dark halo and give rise to the
microlensing events seen by MACHO, EROS and OGLE. However,
Bahcall et al 1994, using the wide field camera of the repaired Hubble
Space Telescope (HST) have searched for red subdwarfs in our MW galaxy
and found very few such stars and that they can contribute no more
than 6\% to the mass of the MW dark halo and no more than 15\%
to the mass of the MW disk.
Thus, the microlensing and HST observations
suggest that most of the MW halo consists
of non baryonic dark matter.

\bigskip
{\bf 3. THE SOLAR NEUTRINO PROBLEM - AN UPDATE}
\smallskip
The Sun is a typical main sequence star that is believed to
generate its energy by fusion of protons into Helium nuclei
through the pp and CNO nuclear reactions chains
which also produce neutrinos. These neutrinos have been detected
on Earth in four pioneering solar neutrino experiments, thus
basically confirming that the sun generates its energy via
fusion of hydrogen into hellium. However, all four experiments
measured solar neutrino fluxes which are
significantly smaller than those predicted by the standard
solar model (SSM) of, e.g., Bahcall and Pinnsenault 1992 (hereafter BP).

The HOMESTAKE  Cl experiment reported (Cleveland et al 1995)
an average production rate of $^{37}$Ar of $2.55\pm 0.25~SNU$
$(1SNU=10^{-36}~s^{-1}$ captures per atom) by solar neutrinos
above the $0.81~MeV$ threshold energy
during 24 years (1970-1993) of observations which is $32\pm 5\%$ of
the $8\pm 3(3\sigma)~SNU$ predicted by the SSM of BP.

KAMIOKANDE II and III observed electron recoils, with energies first
above 9 MeV and later above 7 MeV, from elastic scattering of solar
neutrinos on electrons in water. Their 5.4-year data show a spectrum
consistent with $^8$B solar neutrino flux of (Kajita 1994)
$(2.7\pm 0.2\pm 0.3)\times 10^6 cm^{-2}s^{-1}$, which is
$48\%\pm 9\%$ of that predicted by the SSM of BP.

GALLEX, the European Gallium experiment at the Gran Sasso
underground laboratory, measured (Anselman et al 1995)
a capture rate of solar neutrinos by $^{71}Ga$
of $ 79\pm 10\pm 6~SNU~$
compared with $131.5^{+21}_{-17}~SNU,$ predicted by the SSM
of BP.

SAGE, the Soviet-American Gallium Experiment in the Baksan
underground laboratory reported (Abdurashitov et al 1995) an average
capture rate of solar neutrinos by $^{71}Ga$ of
$74^{+13+5}_{-12-7}~SNU$ during 1990-1993.

The discrepancies between the observations and the SSM predictions
have become known as the solar neutrino problem. Table I
summarize these discrepancies for three different SSM calculations
(Bahcall and Pinsennault 1992; Turck-Chieze and Lopes 1993; Dar
and Shaviv 1994).

$$\matrix{{\rm Exp.}&{\rm Data} &{\rm SSM-BP}& {\rm SSM-TC}&
{\rm SSM-DS}\cr
 {\rm ^{37}Cl~(HOMESTAKE)}& 2.55\pm 0.17\pm 0.18&
8.0\pm 3,0 & 6.4\pm 1.4& 4.2\pm 1.2 \cr
 {\rm ^{71}Ga~(SAGE)~~~~~}& 73^{+18+5}_{-16-7} &
131.5^{+21}_{-17} & 123\pm 7~~ & 113\pm 7~~\cr
 {\rm ^{71}Ga~(GALLEX)~~~~}& 79\pm 10\pm 6 &
131.5^{+21}_{-17} & 123\pm 7~~ & 113\pm 7~~\cr
{\rm H_2O~(KAM~II+III)~}& 2.7\pm 0.2\pm0.3
&5.7\pm 2.5 & 4.4\pm 1.1 & 2.7\pm 0.8 \cr} $$
The results of Homestake, GALLEX and SAGE are in $SNU$, while those
of Kamiokande are in $10^6cm^{-2}s^{-1}.$
Note that the results
of Kamiokande are consistent with the SSM predictions of Dar and
Shaviv (1994; 1995), but the results of the Cl and Gallium experiments
differ significantly from their SSM predictions.
\smallskip
Bahcall and Bethe (1990, 1993) argued that the solution of the solar
neutrino problem requires new physics beyond the Standard Electroweak
Model (Glashow 1961; Weinberg 1967; Salam 1968)
because the signal in the Cl detector due to
the pep, $^7$Be, CNO and $^8$B solar
neutrinos, is smaller than that expected from the $^8$B solar
neutrinos alone as observed by Kamiokande.
But, for a $1.06\times 10^{-42}cm^2$  capture cross section
in $^{37}$Cl of $^8$B neutrinos (Bahcall 1989), the flux observed
by Kamiokande implies a minimal capture rate of $2.86\pm 0.41$ SNU
in the Cl experiment.
During the same period (1986-1993)  Homestake
observed (Cleveland et al. 1995) a capture rate of $2.78\pm 0.35$ SNU
($2.55\pm 0.25~SNU $ is the average over 24 years) which does not provide
conclu
evidence for new physics beyond the standard particle physics model.
However, taken at their face values, the joint results of Kamiokande
and of  Homestake indicate a strong suppression of the SSM $^7$Be flux
(see e.g., Dar 1993, Bahcall 1995), which according to the SSM
is expected to contribute $\sim 1 SNU $ to the capture rate
in the Cl experiment.

Additional indication for the suppression of the
$^7$Be flux is provided by the Gallium experiments:
Since the net reaction in the pp chains and CNO cycle is the
conversion of protons into Helium nuclei, conservation of baryon number,
charge, lepton flavour and energy requires that
  $$ 4p+2e^-\rightarrow {^4He}+2\nu_e+Q~, \eqno\eq $$
where $Q=26.73~MeV,$ i.e.,
two $\nu_e$'s are produced in the Sun
per 26.73 MeV release of nuclear energy. Thus,
if the Sun is approximately in a steady state where its nuclear
energy production rate equals its luminosity (less than 1/2\% of
the solar energy is produced by gravitational contraction)
then the $\nu_{\odot}$ flux at Earth is given by
$$
\phi_{\nu_\odot}= {2L_{\odot}\over Q-2\bar{E}_{\nu}}~{1\over 4\pi D^{2}}
           , \eqno\eq $$

where $L_{\odot}$ is the luminosity of the Sun, D is its distance from
Earth, and $\bar {E}_\nu$ is the average $\nu_\odot$ energy.
The bulk of the solar neutrinos are  pp neutrinos. Consequently,
$\bar {E}_\nu\approx \bar{E}_\nu(pp)\approx 0.265~MeV$,
and Eq.1 yields $\phi_{\nu_\odot}=\Sigma_i\phi_{\nu_i}\approx
6.50\times 10^{10}~ cm^{-2}s^{-1}$. Such a pp flux produces
76 $SNU$ in Gallium.
The tiny flux (relative to the pp flux) of the $^8$B solar neutrinos
observed in Kamiokande increases the signal by 7 $SNU$ to
83 $SNU$. Thus, the $79\pm 10\pm 6$ SNU measured
by GALLEX and the $73^{+18+5}_{-16-7}$ SNU measured by SAGE leave
little room for the $\sim 30 SNU$ contribution of the $^7$Be solar
neutrinos predicted by the SSM
(in the  SSM the flux of $^7$Be neutrinos is $\sim 7\%$ of the flux of pp
 neutrinos but a $^7$Be neutrinos has a capture cross section in
 $^{71}$Ga which is $\sim$ 6.2 times larger than that of a pp neutrino).
\smallskip
Does the Solar Neutrino Problem imply new physics beyond the
standard particle physics model ?

A recent milestone experiment by GALLEX, namely, the calibration
of the GALLEX experiment with an artificial $^{51}$Cr source
(Anselmann et al 1995a) has
eliminated the trivial solution to the Solar Neutrino Problem,
namely, that the accuracy of the results of the radiochemical experiments
has been largely overestimated (the measured
ratio of the production rate
of $^{71}$Ge by neutrinos from an artificial $^{51}$Cr source placed
inside the GALLEX detector to the rate expected from the known source
activity was $R=1.04\pm 0.12$). Standard physics solutions to the
solar neutrino problem have now the difficult task
of explaining the strong  suppression of
the $^7$Be solar neutrino flux. Such a suppression is not ruled out
by standard physics. For instance, collective plasma effects
near the center of the Sun may align the electron and $^7$Be spins
and may change the branching ratios of $e^-$ capture by $^7$Be to
the ground and excited states of $^7$Li. If it causes a
strong reduction in the flux of 0.862 MeV
$^7$Be solar neutrinos with an increase in the flux of 0.384 MeV $^7$Be
solar neutrinos it may explain the observations since
the 0.384 MeV neutrinos are below the threshold for capture in the Cl
detector and have a smaller capture cross section in Gallium.

A more elegant and exciting solution to the Solar neutrino problem
is neutrino oscillations and the MSW effect (Mikheyev and Smirnov 1986;
Wolfenstein 1978,1979). Fig. 4 shows the regions of
mixing parameters $\Delta m^2$ and $sin^2\theta$ of $\nu_e$'s
which can solve the Solar Neutrino Problem. Only future
solar neutrino experiments,
such as the SNO heavy water experiment (Ewan et al 1987)
which will be  able to
detect the conversion of solar $\nu_e$'s into $\nu_\mu$'s and/or
$\nu_\tau$'s, or Super Kamiokande (Kajita 1994) which will be able to
detect deviations from the normal beta decay
energy spectrum of $^7$Be neutrinos,
will be able to confirm that the solution to the Solar Neutrino Problem
requires physics beyond the standard model and that Nature made use
of the beautiful MSW effect.
\smallskip
{\bf 4. COSMIC RAY EVIDENCE FOR NEW PHYSICS ?}
\smallskip
Cosmic ray observations have often been proposed
as evidence for new physics. In most of these cases cosmic ray
puzzles turned out to be the results of a combination of poor
statistics, improperly understood detection techniques and
complicated physics. There are, however, some cosmic
ray anomalies which perhaps require new physics. Here I will
shortly discuss two of them, the atmospheric neutrino anomaly
and the observations of ultrahigh energy cosmic rays above the
Greisen-Zatsepin-Kuzmin energy cutoff.
\smallskip
{\bf 4.1 THE ATMOSPHERIC NEUTRINO ANOMALY}
\smallskip
Atmospheric neutrinos arise from the decay of secondaries ($\pi$, K
and $\mu$) produced in cosmic ray initiated cascades in the atmosphere.
Neutrinos with energies below $\sim 1$ GeV are mainly produced by
$\pi\rightarrow \mu\nu_\mu$ and $\mu\rightarrow e\nu_e\nu_\mu$ decays
and one roughly expects $(\nu_\mu+\bar{\nu}_\mu)/(\nu_e+\bar{\nu}_e)
\approx 2.$ Above $\sim 1$ GeV some of the muons reach the ground before
decay and the $(\nu_\mu+\bar{\nu}_\mu)/\nu_e+\bar{\nu}_e)$ ratio
increases with increasing $E_\nu$. Since the probability of
muon to decay before reaching the ground depends on zenith-angle
the neutrino flavour ratio also depends on zenith-angle. This ratio
has been measured in various large underground neutrino detectors,
NUSEX (Aglietta et al 1989), FREJUS (Berger et al 1990),
SOUDAN-2 (Goodman 1995) IMB-3 (Becker-Szendy et al 1992; 1995)
and KAMIOKANDE (Kajita 1994)
and was compared with Monte Carlo calculations. Some experiments
(SOUDAN-2, IMB-3 and KAMIOKANDE)
found significant discrepancies between the observed and calculated
ratio. This is summarized in Table II.
$$\matrix{{\rm Exp.}&\nu_e &\nu_\mu& \nu_e (MC)& \nu_\mu (MC)&
(\nu_e/\nu_\mu)_{DATA}/(\nu_e/\nu_\mu)_{MC}\cr
{\rm NUSEX~~~~}& 18& 32 & 20.5& 36.8&0.99^{+0.35}_{-0.25}\pm ? \cr
{\rm FREJUS~~~}& 57&108 & 70.6&125.8&1.06^{+0.19}_{-0.16}\pm 0.15 \cr
{\rm SOUDAN-2~}&35.3&33.5& 28.7&42.1&0.64\pm 0.17\pm 0.09 \cr
{\rm IMB-3~~~~}&325&182&257.3&268.0&0.54\pm 0.05\pm 0.12 \cr
{\rm KAM(\leq GeV)}&248&234 &227.6&356.8&0.60^{+0.06}_{-0.05}\pm 0.05 \cr
{\rm KAM(\geq GeV)}&98&135&66.5&162.2&0.57^{+0.08}_{-0.07}\pm 0.07 \cr}$$
In estimating the event rate, NUSEX, FREJUS and SOUDAN-2 used
the flux calculations of Barr et al 1989 while IBM-3 used that of
Lee and Kho 1990 and KAMIOKANDE used that of Honda et al 1990.
If the
$\nu_e$ excess and and the
$\nu_\mu$ deficiency are real they may be due
$\nu_e\leftrightarrow\nu_\mu$
oscillations. The region of neutrino oscillation parameters
($\Delta m^2$ and $sin^2\theta$), which can explain the results
of Kamiokande, is shown in Fig.3 (borrowed from Kajita 1994).
Note that this region for $\nu_e\leftrightarrow \nu_\mu$ oscillations
does not overlap with
that inferred from the MSW solution to the Solar
Neutrino Problem.

It should be noted that the evidence for the atmospheric
neutrino anomaly comes mainly from the light water Cerenkov
detectors. Much larger statistics will be provided by Super Kamiokande
in a couple of years and the long baseline neutrino oscillation
experiments should provide more defenite experimental evidence.
\smallskip
{\bf 4.2 VERY ENERGETIC COSMIC RAYS BEYOND THE GZK CUTOFF}
\smallskip
Greisen (1966) and Zatsepin and Kuzmin (1966) have pointed out that
if very high energy cosmic rays are produced at cosmological distances,
as inferred from their isotropy, their energy spectrum should be cutoff
around $E\sim m_\pi(2m_p+m_\pi)/4E_\gamma\sim 10^{20}eV$, the
threshold energy for photoproduction in head-on collisions with
photons of the microwave background radiation (MBR) whose average energy
is $\bar{E}_\gamma\approx 2.7kT\approx 6.32\times 10^{-4}eV$. For
protons above $3\times 10^{20}eV$ the attenuation length is less than
30 $Mpc$ (Stecker 1968; Hill and Schramm 1985; Yoshida and Teshima
1993). Nuclei and gamma rays have even shorter attenuation lengths
(Puget et al 1976; Wdowczyk et al 1972).

By combining all the data accumulated
for more than 30 years on the the highest energy cosmic rays
from the extensive air shower array experiments at Volcano
Ranch, Haverah Park, Sydney, Yakutsk, Dugway and Akeno significant
evidence for the GZK has been accumulated.
That is, only several cosmic rays exceeding $10^{20}eV$
have been observed compared with expectation of more than 25
if there is no cutoff and their energy extends beyond $10^{20}eV$
with the same power index (Hayashida et al 1994).

Recently, however, the two most energetic cosmic rays have
been observed by the Fly's Eye ($E=3.2\pm0.9\times 10^{20}eV$;
 Bird et al 1994), by the Akeno Giant Air Shower Array
($E=(1.7-2.6)\times 10^{20}eV$; Hayashida et al 1994)
from directions in the sky where no nearby cosmic accelerators,
such as active galactic nuclei, have been seen (because of their high
magnetic rigidity the arrival directions of these cosmic rays should
point aproximately to their sources). It suggests a diffuse origin
of the ultrahigh energy cosmic rays. But, what can be this origin?
The lack of known conventional diffuse sources of ultrahigh energy cosmic
rays calls for alternative diffuse sources such as cosmic strings or
annihilation of magnetice monopoles or of other very massive relic
particles from the Big-Bang.

A detailed discussion of the possible nature and origin
of the ultrahigh energy cosmic rays
has been made by Elbert and Sommers 1994. However,
many more cosmic ray
events above the GZK cutoff are needed before any definite
conclusions regarding their identity and origin can be made.
\endpage
\centerline {{\bf REFERENCES}}
\smallskip

Abdurashitov, J.N. et al., 1995, Nucl. Phys. B (Proc. Suppl.)
{\bf 38}, 60

Aglietta, M. et al, 1989, Europhys. Lett. {\bf 8}, 611

Anselmann, P. et al., 1995, Nucl. Phys. B (Proc. Suppl.) {\bf 38}, 68

Anselmann, P. et al., 1995a, Phys. Lett. B (in press).

Alcock, C. et al., 1993, Nature, {\bf 365 }, 621

Alcock, C. et al., 1995, Univ. of Washington Preprint, Jan. 1995

Aubourg, E. et al., 1993, Nature, {\bf 365}, 623

Balser, D.S., Bania, T.M., Brockway, C.J., Rood, R.
T. and Wilson, T.L., 1994, ApJ

Bahcall, J.N., 1989, {\it Neutrino Astrophysics} (Cambridge Univ. Press,
1989)

Bahcall, J.N. et al., 1994, ApJ. {\bf 435}, L51

Bahcall, J.N., 1995, Nucl. Phys. B (Proc. Suppl.) {\bf 38}, 98

Bahcall, J.N. and Bethe, H.A. 1990, Phys. Rev. Lett. {\bf 65}, 2233

Bahcall, J.N. and Bethe, H.A. 1993, Phys. Rev. D. {\bf 47}, 1298.

Bahcall, J.N. and Pinsonneault, M., 1992, Rev. Mod. Phys. {\bf 64}, 885.

Barr, G. et al., 1989, Phys. Rev. D {\bf 39} 3532

Becker-Szendy, R. et al., 1992, Phys. Rev. D {\bf 46}, 3720

Becker-Szendy, R. et al., 1995, Nucl. Phys. B (Proc. Suppl.)
{\bf 38}, 333

Berger, Ch.  et al., 1990, Phys. Lett. {\bf 245}, 305.

Bethe, H., 1939, Phys. Rev. {\bf 55}, 103.

Bird, D.J., et al., 1994 ApJ. {\bf 424}, 491

Blandford, R.D. et al., 1989, Science {\bf 245}, 824

Blandford, R.D. and Narayan, R., 1992, Ann. Rev. Astron.
     Astrophys. {\bf 30}, 311.

Cholson, O., Astron. 1924, Nachr. {\bf 221}, 329

Carswell, R.F. et al., 1994,  MNRAS, in press

Caughlan, G.R. and Fowler, W.A., 1988, Atom. and Nucl. Data Tables,
{\bf 40}, 284

Clayton, D., 1968, {\it Principles of Stellar Evolution \&
Nucleosynthesis} (McGraw-Hill)

Cleveland et al., 1995 Nucl. Phys. B (Proc. Suppl.) {\bf 38}, 47

Damour, T. and Taylor, J.H., 1991, ApJ. {\bf 366}, 501

Dar, A., 1992, Nucl. Phys. B (Proc. Suppl.) {\b 28A}, 321top

Dar, A., 1993, in {\it Particle Physics and Cosmology} (eds.
V. Matveev et al.)

Dar, A., 1993a, Proc. 1st Rencontre De Vietnam on Astroparticle
Phys. (ed. Tran Thanh Van)

Dar, A., Goldberg, J. and Rudzsky, M. 1992, Technion Preprint
PHR- 92-12

Dar, A. and Shaviv, G., 1994, Proc. VI Int. Workshop on
Neutrino Telescopes, p. 303  (ed. M. Baldo-Ceolin); Shaviv, G.,
1995, Nucl. Phys. B (Proc. Suppl.) {\bf 38}, 81.

Einstein, A., 1936, Science {\bf 84}, 506

Ewan, G.T. et al., 1987, Sudbury Neutrino Observatory Proposal
SNO-87-12

Elbert, J.W.  and Sommers, P., 1994 (to be published)

Freedman, W.L. et al., 1994, Nature, {\bf 371}, 757

Geiss, J., 1993, in {\it Origin and Evolution of the Elements}, eds.
N. Prantzos et al. (Cambridge University Press,
1993) p. 89

Glashow, S. 1961, Nucl. Phys. {\bf 22}, 579

Goodman, M. C., 1995, Nucl. Phys. B (Proc. Suppl.) {\bf 38}, 337

Gould, A., 1995, Nucl. Phys. B (Proc. Suppl.) {\bf 38}, 371

Greisen, K., 1966, Phys. Rev. Lett. {\bf 16}, 748

Grossman, S.A. and Narayan, R., 1988, ApJ. Lett. {\bf 324}, L37

Hayashida N. et al., 1994, Phys. Rev. Lett. {\bf 73}, 3491

Hill, C.T. and Schramm, D,N., 1985, Phys. Rev. D {\bf D31}, 564

Hogan, C., 1994, Astrophys. Bull. Board preprint  astro-ph/9407038

Honda, M. et al., 1990, Phys. Lett. B {\bf 248}, 193

Izotov, Y., Thuan. X.T. and Lipovetsky, V.A., 1994, ApJ. {\bf 435}, 647

Kajita, T., 1994 ICCR-Report 332-94-27 (December 1994).

Kolb, R. and Turner, M., 1991, {\it The Early Universe} (Addison
Wesley-1990)

Langston, G.I. et al., 1989, Astr. J. {\bf 97}, 1283

Langston, G.I.  et al., 1990, Nature {\bf 344}, 43 (1990)

Lee, H. and Koh, Y.S., 1990, Nuov. Cim. {\bf 105B}, 883.

Linsky, J.L. et al., 1993, ApJ, {\bf 402}, 694

Loveday, J. et al., 1992, ApJ. {\bf 390}, 338

Lynds, R. and Petrosian, V., 1989, ApJ. {\bf 336}, 1

Mana, C. and Martinez, M., 1993, Nucl. Phys. B (Proc. Suppl.)
{\bf 31}, 163

Mannheim, P.D. and Kazanas, D. 1989
    Ap. J. {\bf 342}, 635

Mather, J.C. et al., 1994,  ApJ, {\bf 420}, 439

Mathews, G.J. et al., 1993, ApJ,  {\bf 403}, 65

Mathews, G.J., and Malaney, R.A., 1993, Phys. Rep., {\bf 229}, 147

Mikheyev, P. and Smirnov, A. Yu. 1986, Nuov. Cim. {\bf 9C}, 1986

Milgrom, M. and  Bekenstein, J., 1984, ApJ. {\bf 286}, 7

Moscoso, L., 1995, Nucl. Phys. B (Proc. Suppl.) {\bf 38}, 387

Paczynski, B., 1986, Astrophys. J. {\bf 304}, 1

Pagel, E.J. et al., 1992, MNRAS, {\bf 255}, 325

Particle Data Group, 1994,  Phys. Rev. {\bf D50}, 1173

Peebles, P.J.E., 1966, ApJ., {\bf 146}, 542

Peebles, P.J.E., 1993, {\it Principles of Physical Cosmology},
(Princeton Series in Physics)

Pelt, J. et al., 1994, Astron. and Astrophys. {\bf 286}, 775

Puget, J.L. et al., 1976, ApJ. {\bf 205}, 638

Rix, H.W., Schneider, D.P. and Bahcall, J.N., 1992,  Astron. J.

Rood, R.T., Bania, T.M., and   Wilson, T.L. 1992, Nature,
{\bf 355}, 618

Sackett, P. et al., 1994, Nature {\bf 370}, 441

Salam, A. 1968, in "Elementary Particle Theory", p. 367 (ed.
N. Svartholm,  Almqvist and Wiksells, Stockholm 1968)

Sanders,  R.H., Asstron. Astrophys. 1984, Lett. {\bf 136}, L21

Sanders, R.H., 1990, Astron. Astrophys. Rev. {\bf 2}, 1
and references therein.

Schild, R.E., 1990, Astron. J. {\bf 100}, 1771

Skillman E.D. and Kennicutt, R.C. Jr., 1993, ApJ, {\bf 411}, 655

Smith, V.V., Lambert, D.L. and Nissen, P.E., 1993, ApJ,
 {\bf 408}, 262

Smith, M.S., Kawano, L.H. and Malaney, R.A., 1993, ApJ (Suppl.),
{\bf 85}, 219

Songaila, A. et al., 1994, Nature, {\bf 368}, 599

Soucail, G. et al., 1988, Astro. Astrophys. Lett. {\bf 172}, L14

Spite, M. and Spite, F., 1982a, Astron. and Astrophys., {\bf 115}, 357

Spite, M. and Spite, F., 1982b, Nature, {\bf 297}, 483

Stecker, F.W., 1968, Phys. Rev. Lett. {\bf 21}  1016

Sutherland, W. et al., 1995, Nucl. Phys. B (Proc. Suppl.) {\b 38}, 380

Thorburn, J.A., 1994, ApJ, {\bf 421} 318

Tremaine, S. and Gunn, J.E. 1979, Phys. Rev. Lett. {\bf 42}, 407

Turck-Chieze, S. et al. 1988, Ap. J. {\bf 335}, 415

Turck-Chieze, S. and Lopes, I. 1993, Ap. J. {\bf 408}, 347

Tyson J. A. et al., 1990, ApJ, {\bf 349}, L1

Tyson, J.A., 1991, A.I.P. Conf. Proc. {\bf 222}, 437

Udalski, A., et al., 1993, Acta Astronomica, {\bf 43}, 289

Vanderriest, C.  et al., 1989, Astron. Astrophys. {\bf 215} 1

Wagoner, R.V., Fowler, W.A. and Hoyle, F., 1967, ApJ, {\bf 148}, 3

Wagoner, R.V. 1973, ApJ, {\bf 179}, 343

Walker, T.P. et al., 1991, ApJ, {\bf 376}, 51

Webb, J.K., 1991,  MNRAS, {\bf 250}, 657

Weinberg, S. 1967, Phys. Rev. Lett. {\bf 19}, 1264

Weinberg, S. 1972, {\it Gravitation And Cosmology} (John
Wiley \& Sons 1972).

White, S.D.M. et al., 1993, Nature {\bf 366}, 429

Will, C.M., 1984, Physics Reports {\bf 113}, 345 and references
   therein.

Wilson, T.L. and Rood, R.T., 1994, Ann. Rev. Astron. Astrophys.
{\bf 32}, in press

Wdowczyk, J. et al., 1972,  J. Phys. A {\bf 5}, 409

Wolfenstein, L. 1978, Phys. Rev. {\bf D17}, 2369

Wolfenstein, L. 1979, Phys. Rev. {bf D20}, 2634

Yang, J. et al., 1984, ApJ. {\bf 281}, 493

Yoshida, S. and Teshima, 1993, M., Prog. Theor. Phys. {\bf 89}, 833

Zatsepin , G.T. and Kuzmin, V.A., 1966, Pis'ma Zh. Eksp. Ther. Fiz.
{\bf 4}, 1966

Zwicky, F., 1933, Helv. Phys. Acta, {\bf 6}, 110

\endpage

\centerline{{\bf FIGURE CAPTIONS}}
\smallskip
{\bf Fig. 1}: The ratios between the EGR prediction and the observations
of the deflection and time delay of light from distant quasars and
galaxies by galaxies or clusters of galaxies, displayed
at the impact parameter of the deflected light relative to the center
of the lens, for the Einstein Cross Q2237+05,
the Einstein Ring MG1654+1346, the double quasar Q0957+561 and
the Einstein Arcs in A370, Cl2244-02, and Cl0024+1654.
The estimated errors in the ratios include the quoted observational
errors and the errors in the theoretical estimates due only to
errors in measured parameters and
the absence of precise knowledge of $\Omega$ and $h~$, but not
systematic errors.
\smallskip
{\bf Figure 2.} (a) The primordial mass fraction of $^4$He and
the abundances (by numbers) of D, $^3$He and $^7$Li
as a function of $\eta_{10}$ as predicted by SBBN theory.
Also shown are their observed values extrapolated to zero age,
as summarized in section III.
The vertical line indicates the value $\eta_{10}=1.6$.
(b) The values of $\chi^2$ (left scale)
and the corresponding confidence level (right scale)
of the agreement between the predicted abundances and those
inferred from observations, as function of $\eta_{10}~.$
Best agreement is obtained for $\eta_{10}\approx 1.60~$ with
a confidence level above 70\%.
\smallskip
{\bf Figure 3.} Likelihood contours for the MACHO mass
derived by the MACHO collaboration from their  3 microlensing
events of LMC stars. A spherical MW halo of equal mass Machos
was assumed (from Alcock et al 1995).

\smallskip
{\bf Figure 4.} The allowed parameter regions of the MSW solution
to the solar neutrino problem adopting the SSM predictions of
Bahcall and Pinsennault and of Turck-Chieze and Lopes (from Kajita
1994).
\smallskip
{\bf Figure 5.} The allowed parameter regions of the neutrino
oscillation solutions to the atmospheric neutrino anomaly seen
by Kamiokande, IMB-3  and Soudan-2 (from Kajita 1994).

\endpage
\end